\documentclass[%
aps,
pre,
amsmath,amssymb,
reprint,
]{revtex4-2}

\usepackage{xcolor}
\usepackage{graphicx}
\usepackage[normalem]{ulem}
\newcommand{\rev}[1]{#1}

\begin{document}
	
	\title{First Passage Times for Variable-Order Time-Fractional Diffusion}
	\author{Wancheng Li}
	\affiliation{School of Mathematics and Statistics, UNSW, Sydney, Australia}
	\author{Daniel S. Han}
	\email{daniel.han@unsw.edu.au}
	\affiliation{School of Mathematics and Statistics, UNSW, Sydney, Australia}
	\date{\today}
	
	\begin{abstract}
		We derive the asymptotic first passage time (FPT) distribution for space-dependent variable-order time-fractional diffusion, where the fractional exponent $\alpha(x)$ varies with position.
		On a bounded interval with an absorbing and a reflecting boundary, we show that the survival probability decays as $\Psi(t)\sim C\,t^{-\alpha_*}/[\ln (t/\tau)]^{\nu}$, where $\alpha_*$ is the minimum value of the fractional exponent and $\nu$ is determined by the location and shape of the minimum.
		The exponent $\nu$ is fixed by the location and order of the minimum, taking the value $1/k$ at a $k$\textsuperscript{th}-order interior minimum, $1$ at the reflecting boundary and $2$ at the absorbing boundary.
		For a constant fractional exponent, $\nu=0$, so the logarithmic factor distinguishes a continuously varying order field from a homogeneous medium.
		We recover both exponents from simulated first passage times by a direct linear fit, with no amplitude or geometric input.
		We validate the theory against exact Laplace-space solutions and Monte Carlo simulations for linear and nonlinear profiles of $\alpha(x)$.
	\end{abstract}
	
	\maketitle

\section{Introduction}
\label{sec:intro}

	Anomalous diffusion, characterized by a mean square displacement (MSD) $\langle x^2 (t) \rangle \sim t^{\alpha}$ with fractional exponent $\alpha$, underlies transport in a broad range of physical \cite{tiedje1981physical,uchauikin2013fractional,baggioli2021anomalous,sposini2022towards,caucal2022anomalous,bodrova2025anomalous,romano2026anomalous}, chemical \cite{liu2022heterogeneous,avula2023understanding,rajyaguru2024diffusion,rajyaguru2025quantifying}, and biological systems \cite{tolic2004anomalous,hofling2013anomalous,fedotov2018memory,burrage2024fractional,vilk2025strong}.
	A comprehensive review of anomalous diffusion for constant fractional exponent $\alpha$ has been presented by Metzler and Klafter \cite{metzler2000random}.
	Recent experimental evidence in biology \cite{waigh2023heterogeneous} for intracellular vesicles \cite{han2020deciphering,fedotov2021variable,korabel2021local}, cell migration \cite{korabel2022hemocytes} and transport in amorphous semi-conductors \cite{sibatov2024variable} have driven developments surrounding variable-order fractional diffusion equations.
	Variable-order fractional diffusion is when the fractional exponent varies depending on space and/or time, $\alpha \rightarrow \alpha(x,t)$ \cite{lorenzo2002variable,sun2009variable}.
	This dependence arises from the need to model anomalous diffusion in environments where the medium is inhomogeneous.

	An established model for anomalous diffusion on a finite domain $[0,L]$, in inhomogeneous media is the space-dependent variable-order time-fractional diffusion equation \cite{chechkin2005fractional,sun2009variable, korabel2010paradoxes, fedotov2012subdiffusive, straka2018variable}
	\begin{equation}
		\partial_t p(x,t)
		=
		\partial^2_x
		\left[
		D_{\alpha(x)}
		\mathcal{D}_t^{1-\alpha(x)} p(x,t)
		\right],
		\label{eq:vofde}
	\end{equation}
	where $p(x,t)$ is the probability density function (PDF) of a particle at position $x\in[0,L]$ and time $t$.
	The fractional exponent is $\alpha(x)$ and the fractional diffusion coefficient is $D_{\alpha(x)}=a^2/(2 t_0 \tau^{\alpha(x)})$ with a time scale $\tau$, a scaling parameter $t_0$ and a length scale $a$. 
	The need for both the scaling parameter and time scale \rev{arises due to the diffusion limit and} has been extensively discussed in \cite{angstmann2026global} and, for convenience, we set $t_0 = 1$ from now on.
 \rev{Note that a space-dependent fractional exponent is a model choice and allowing the generalized diffusion coefficient to vary with position is an alternative.}
	The Riemann-Liouville derivative $\mathcal{D}_t^{\,1-\alpha(x)}$ with $0<\alpha(x)<1$ is defined by
	\begin{equation}
		\mathcal{D}_t^{\,1-\alpha(x)} p(x,t)
		:=
		\frac{1}{\Gamma(\alpha(x))}\partial_t\int_0^t \frac{p(x,t')}{(t-t')^{1-\alpha(x)}}\,dt',
		\label{eq:RL_def}
	\end{equation}
	where $\Gamma(\cdot)$ is the Gamma function. 
	We denote the Laplace transform of $f(t)$ by $\hat{f}(s) := \mathcal{L}\{f(t)\} = \int_0^\infty e^{-st} f(t)\,dt$.
	The Laplace transform of \eqref{eq:RL_def} satisfies
	\begin{equation}
		\mathcal{L}\left\{\mathcal{D}_t^{\,\beta} f(t)\right\}
		=
		s^{\beta}\,\hat{f}(s)
		-
		\left.\mathcal I_t^{\,1-\beta}f(t)\right|_{t=0^+},
		\label{eq:RL_LT_general}
	\end{equation}
	where $\mathcal I_t^{\,1-\beta}$ denotes the fractional integral of order $1-\beta$.
	For functions that are locally integrable and bounded near $t=0$, the fractional integral term vanishes, and \eqref{eq:RL_LT_general}
	reduces to
	\begin{equation}
		\mathcal{L}\left\{\mathcal{D}_t^{\,\beta} f(t)\right\}
		=
		s^{\beta}\,\hat{f}(s).
		\label{eq:RL_LT}
	\end{equation}
	Taking the Laplace transform, \eqref{eq:vofde} becomes
	\begin{equation}
		s\hat{p}(x,s) - p(x,0) = \partial_x^2\left[ D_{\alpha(x)} s^{1-\alpha(x)} \hat{p}(x,s) \right].
        \label{eq:vofde_laplace}
	\end{equation}

 \rev{Equation \eqref{eq:vofde} is one of several inequivalent ways of making the fractional exponent space dependent, and the equation alone does not determine which arrangement of the operators is meant \cite{chechkin2005fractional,straka2018variable,fedotov2012subdiffusive}.
    Section~\ref{sec:ctrw} therefore derives \eqref{eq:vofde} from a continuous time random walk (CTRW) with a position-dependent waiting time density, which fixes the ordering and shows why the diffusion limit must be taken globally rather than at each point.}

	Since the derivation of \eqref{eq:vofde_laplace} \cite{chechkin2005fractional}, much progress has been made on numerical \cite{chen2010numerical} and analytical \cite{chechkin2005fractional,korabel2010paradoxes,fedotov2019asymptotic,roth2020inhomogeneous} solution approximations.
	Notably, the asymptotic solution \cite{fedotov2019asymptotic} and the exact Laplace space solution \cite{roth2020inhomogeneous} have been found for the case when the fractional exponent is a linear function of space, $\alpha(x) = c+bx$.
	In recent work, the PDF for variable-order time-fractional diffusion has been fit using brute force methods \cite{fedotov2021variable} and MSD calculations have been performed using Monte Carlo simulations \cite{sibatov2024variable}.
	Even with these methods, it remains difficult to distinguish when variable-order fractional diffusion exists in experimental systems and to infer the parameters of a space varying fractional exponent given data.
    This is due to the lack of theoretical predictions for statistical properties that can be compared with experiments.
 \rev{To our knowledge, no scaling prediction for the first passage time (FPT) distribution has previously been available for variable-order time-fractional diffusion.}
	Our results below provide a way to quantitatively test experimental observations for variable-order fractional diffusion.
	
	In this paper, we formulate the FPT problem for the variable-order time-fractional diffusion equation for any twice differentiable fractional exponent, $\alpha(x)$.
	We find that the FPT distribution scales asymptotically as a power law dependent on the minimum value of $\alpha(x)$ with a logarithmic correction dependent on the location and order of the minimum.
    We illustrate the general result with two examples, $\alpha(x) = c+bx$ and $\alpha(x)=\alpha_{0}- A(\phi(x-x_1) + \phi(x-x_2))$ where $ \phi (x) = \exp ({-x^2/(2\sigma^2)})$, confirming the theory using numerical inverse Laplace transforms and Monte Carlo simulations.
	
	%%%%%%%%%%%%%%%%%%%%%%%%%%%%%%%%%%%%%%%%%%%%%%%%%%%
	%%%%%%%%%%%%%%%%%%%%%%%%%%%%%%%%%%%%%%%%%%%%%%%%%%%
	
\section{The underlying continuous time random walk}
\label{sec:ctrw}
    
    Equation \eqref{eq:vofde} is one of several ways of making the fractional exponent space dependent, with three arrangements $\partial^2_x[D_{\alpha(x)}\mathcal{D}_t^{1-\alpha(x)}p]$, $\mathcal{D}_t^{1-\alpha(x)}\partial^2_x p$ and $\partial_x[D_{\alpha(x)}\mathcal{D}_t^{1-\alpha(x)}\partial_x p]$ describing different microscopic rules and different long-time behaviour \cite{chechkin2005fractional,straka2018variable}.
    Fedotov and Falconer \cite{fedotov2012subdiffusive} showed that the associated master equation is structurally unstable, so an arbitrarily small spatial variation of $\alpha$ alters the long-time behaviour qualitatively rather than perturbatively.
    The derivation of this master equation from the limit of an underlying stochastic process already exists \cite{angstmann2026global} but we summarize it here for completeness.
    
    Consider a walker on a one-dimensional lattice of spacing $a$ making unbiased nearest-neighbour jumps, waiting at each site for a time drawn from a position-dependent density $\psi(x,t)$ with survival function $\Psi(x,t)=1-\int_0^t\psi(x,t')\,\mathrm{d}t'$.
    The generalized master equation is
    \begin{equation}
        \begin{split}
            \frac{\partial P(x,t)}{\partial t}
            ={}&\frac12\int_0^t\! K(x-a,t-t')P(x-a,t')\,\mathrm{d}t'\\
            &+\frac12\int_0^t\! K(x+a,t-t')P(x+a,t')\,\mathrm{d}t'\\
            &-\int_0^t\! K(x,t-t')P(x,t')\,\mathrm{d}t' ,
        \end{split}
        \label{eq:ctrw-master}
    \end{equation}
    where $P(x,t)$ is the probability of finding the walker at $x$ at time $t$ and the memory kernel is fixed in Laplace space by
    \begin{equation}
        \hat K(x,s)=\frac{\hat\psi(x,s)}{\hat\Psi(x,s)} .
        \label{eq:ctrw-kernel}
    \end{equation}
   
    We choose a heavy-tailed waiting time, the Mittag-Leffler density, with a position-dependent exponent rescaled such that a global diffusion limit can be taken
    \begin{equation}
        \psi(x,t)=\frac{t^{\alpha(x)-1}}{\tau\,t_0^{\alpha(x)}}
        E_{\alpha(x),\alpha(x)}\!\left(-\frac{1}{\tau}\left(\frac{t}{t_0}\right)^{\alpha(x)}\right),
        \label{eq:ml-rescaled}
    \end{equation}
    where $\tau$ is a dimensionless scale factor, $t_0$ is the characteristic time and the kernel in Laplace space is $\hat K(x,s)=s^{1-\alpha(x)}/(t_0\tau^{\alpha(x)})$.
    The generalized diffusion coefficient is
    \begin{equation}
        D_{\alpha(x)}
        =\lim_{\substack{a\to0\\ \tau \to0}}\frac{a^{2}}{2\tau}\,t_0^{-\alpha(x)}.
        \label{eq:global-limit}
    \end{equation}
    This global diffusion limit is equivalent to the CTRW derivation presented in \cite{straka2018variable}.
    Taking the global diffusion limit of \eqref{eq:ctrw-master} with the Mittag-Leffler density \eqref{eq:ml-rescaled} then yields \eqref{eq:vofde}.
    Equation \eqref{eq:vofde} can be derived from a discrete time random walk with Sibuya-distributed waiting times and also for higher dimensions \cite{angstmann2026global}.
    The corresponding equation for flowing media where a velocity field is added to the same class of random walk is derived in \cite{nie2025analysis}.

	%%%%%%%%%%%%%%%%%%%%%%%%%%%%%%%%%%%%%%%%%%%%%%%%%%%
	%%%%%%%%%%%%%%%%%%%%%%%%%%%%%%%%%%%%%%%%%%%%%%%%%%%
	
\section{First passage times in one dimension}
\label{sec:fpt1d}
    We consider the FPT distribution of the random walk whose governing equation is \eqref{eq:vofde} for general $\alpha(x)$ given the initial condition 
    \begin{equation}
        p(x,0) = \delta(x-x_0),
        \label{eq:ICFPT}
    \end{equation}
    for $x \in (0,L]$.
	The FPT distribution, or the survival probability, is calculated via \cite{redner2001guide}
	\begin{equation}
		\Psi(t)=\int_0^L p(x,t)\,dx,
		\label{eq:SurvivalProbability}
	\end{equation}
	and the FPT PDF is obtained from $\psi(t)=-d\Psi(t)/dt$,
	where $p(x,t)$ is the PDF of the random walk governed by \eqref{eq:vofde} given initial condition \eqref{eq:ICFPT}.
    After taking the Laplace transform of \eqref{eq:vofde}, the density $\hat{p}(x,s)$ is governed by \eqref{eq:vofde_laplace} with the absorbing boundary condition
    \begin{equation}
        \hat p(0,s)=0,
    \end{equation} and the reflecting boundary condition
    \begin{equation}
        \partial_x\!\Big(D_{\alpha(x)}s^{1-\alpha(x)}\hat p(x,s)\Big)\Big|_{x=L}=0.
    \end{equation}
    If we define
    \begin{equation}
        F(x,s):=D_{\alpha(x)}s^{1-\alpha(x)}\hat p(x,s),
        \label{eq:Fs-def}
    \end{equation}
    then $\hat p(x,s)=s^{\alpha(x)-1}F(x,s)/D_{\alpha(x)}$, and \eqref{eq:vofde_laplace} yields
    \begin{equation}
        \label{eq:Fs-eqn}
        -F''(x,s)+q(x,s)F(x,s)=\delta(x-x_0),
    \end{equation}
    with 
    \begin{equation}
         q(x,s)={s^{\alpha(x)}}/{D_{\alpha(x)}}={2}(\tau s)^{\alpha(x)} /{a^2},
    \end{equation}
    and boundary conditions $F(0,s)=0,$ and $F'(L,s)=0$.
    To solve \eqref{eq:Fs-eqn} asymptotically as $s\rightarrow 0$, we can write \eqref{eq:Fs-eqn} as
    \begin{equation}
    \label{eq:asy-ode}
       -\lim_{{s \to 0} }F''(x, s)=\delta(x-x_0).
    \end{equation} 
    since $q(x,s) \to 0$ uniformly and $F(x,s)$ is bounded as $s\to 0$.
    The solution of \eqref{eq:asy-ode} is $F(x,0) = \min \{x,x_0\}$ where $x, x_0 \in [0,L]$.
    Now substituting \eqref{eq:Fs-def} into \eqref{eq:SurvivalProbability}, we obtain 
    \begin{equation}
        \hat\Psi(s)=\frac{2\tau}{a^2}\int_0^L (\tau s)^{\alpha(x)-1}F(x,s){d}x.
    \end{equation}
    Since by a comparison argument
    \begin{equation}
        0\le F(x,0)-F(x,s)\le L^2\|q(x,s)\|_{\infty}F(x,0),
    \end{equation}
    where $\|q(x,s)\|_{\infty}:= \displaystyle \sup_{ x}q(x,s)$, and $\displaystyle\lim_{s\to 0} \| q(x,s) \|_\infty =0$, we have
    \begin{equation}
    \begin{split}
        \lim_{s \to 0} & \int_0^L  (\tau s)^{\alpha(x)-1}F(x,s)\,{d}x =\\
        &\left(1+O(\|q(x,s)\|_{\infty})\right)\int_0^L (\tau s)^{\alpha(x)-1}F(x,0)\,{d}x.
    \end{split}
    \end{equation}
    Hence \eqref{eq:SurvivalProbability} in Laplace space becomes 
    \begin{equation}
        \hat\Psi(s)\sim\frac{2\tau}{a^2}\int_0^L (\tau s)^{\alpha(x)-1}F(x,0)\,\mathrm{d}x,
        \label{eq:asy-final}
    \end{equation} as $s \to 0$.
    As long as $\alpha(x)$ is twice differentiable, we can use Laplace's method to approximate \eqref{eq:asy-final}, such that 
    \begin{equation}
        \hat\Psi(s)\sim\frac{2}{sa^2}\int_0^L e^{\alpha(x)\ln(\tau s)}F(x,0)\,{d}x .
    \label{eq:psi_laplace_method}
    \end{equation}
  In what follows, we consider various cases for the fractional exponent, $\alpha(x)$, and provide asymptotic expressions for the FPT distribution. 	

\subsection{Unique interior minimum}
\label{sec:interior}
    If $\alpha(x)$ has a unique interior minimum $\alpha_* = \alpha(x_*)$ at $x_* \in (0,L)$, then using 
    \begin{equation}
        \alpha(x)=\alpha_*+\frac12\alpha''(x_*)(x-x_*)^2+O\big((x-x_*)^3\big),
    \end{equation}
    \eqref{eq:psi_laplace_method} becomes 
    \begin{equation}
    \begin{split}
        \hat\Psi (s) &\sim \frac{s^{\alpha_*-1}}{D_{\alpha_*}} F(x_*,0)\int_{0}^{L}
        e^{-\frac{\alpha''(x_*)}{2}|\ln(\tau s)|(x-x_*)^2} dx \\
        & \sim \frac{s^{\alpha_*-1}}{D_{\alpha_*}}  F(x_*,0)
        \sqrt{\frac{2\pi}{\alpha''(x_*)\,|\ln(\tau s)|}},
    \end{split}
        \label{eq:asy-IM}
    \end{equation}
    where we have used change of variable $u=\sqrt{|\ln(\tau s)|}\,(x-x_*)$ to evaluate the integral about $x_*$.
    Here we have assumed that $\alpha''(x_*) \neq 0$.
    Applying the Tauberian theorem \cite{feller1991introduction} to \eqref{eq:asy-IM} yields
    \begin{equation}
        \Psi(t)\sim \frac{Ct^{-\alpha_*}}{ \sqrt{\ln \left( t/\tau \right)}},
    \label{eq:psi_interior_time}
    \end{equation}
    where $C = \frac{F(x_*,0)}{D_{\alpha_*}\Gamma(1-\alpha_*)}
    \sqrt{\frac{2\pi}{\alpha''(x_*)}}$.

\subsection{Several equal isolated minima}
\label{sec:multi}
If $\alpha(x)$ has $m$ isolated interior minima at $x_1,\ldots,x_m\in(0,L)$, all sharing the same minimum value $\alpha_*$, then using
    \begin{equation}
        \alpha(x)=\alpha_*+\tfrac{1}{2}\alpha''(x_j)(x-x_j)^{2}+O\left((x-x_j)^{3}\right),
    \label{eq:alpha_expand_multi}
    \end{equation}
    around each $x_j$ and $\alpha''(x_j) \neq 0$, \eqref{eq:psi_laplace_method} becomes
    \begin{equation}
    \begin{split}
        \hat\Psi(s) &\sim \frac{s^{\alpha_*-1}}{D_{\alpha_*}}
        \sum_{j=1}^{m}F(x_j,0)\int_{0}^{L}
        e^{-\frac{1}{2}\alpha''(x_j)|\ln(\tau s)|(x-x_j)^{2}}\,dx \\
        &\sim \frac{s^{\alpha_*-1}}{D_{\alpha_*}}
        \sqrt{\frac{2\pi}{|\ln(\tau s)|}}
        \sum_{j=1}^{m}\frac{F(x_j,0)}{\sqrt{\alpha''(x_j)}},
    \end{split}
    \label{eq:psi_multi_laplace}
    \end{equation}
    where we have applied the same Gaussian localization as in (\ref{eq:asy-IM}) within disjoint neighbourhoods of each $x_j$ and summed the leading contributions. 
    Applying the Tauberian theorem to (\ref{eq:psi_multi_laplace}) yields
    \begin{equation}
        \Psi(t)\sim\frac{C\,t^{-\alpha_*}}{\sqrt{\ln\bigl( t/\tau \bigr)}},
    \label{eq:psi_multi_time}
    \end{equation}
    where $C=\frac{1}{D_{\alpha_*}\Gamma(1-\alpha_*)}\sum_{j=1}^{m}F(x_j,0)\sqrt{\frac{2\pi}{\alpha''(x_j)}}$.
    Equation \eqref{eq:psi_multi_time} has the same time scaling as \eqref{eq:psi_interior_time} and only differs by a multiplicative factor due to the presence of multiple minima.
    If the fractional exponent $\alpha(x)$ contains multiple minima with different values, only the lowest survives in the asymptotic limit.
    
\subsection{Higher order minimum}
\label{sec:korder}
    More generally, if the first nonzero derivative of $\alpha$ at $x_*$ is of even order $k$, so that $\alpha'(x_*) = \cdots = \alpha^{(k-1)}(x_*) = 0$ and $\alpha^{(k)}(x_*) > 0$, then using
    \begin{equation}
        \alpha(x) = \alpha_* + \frac{\alpha^{(k)}(x_*)}{k!}(x-x_*)^k + O\big((x-x_*)^{k+1}\big),
        \label{eq:alpha_expand_k}
    \end{equation}
    \eqref{eq:psi_laplace_method} becomes
    \begin{equation}
    \begin{split}
        \hat\Psi(s) &\sim \frac{s^{\alpha_*-1}}{D_{\alpha_*}}F(x_*,0)\int_0^L e^{-\frac{\alpha^{(k)}(x_*)}{k!}|\ln(\tau s)|(x-x_*)^k}\,dx\\
        &\sim \frac{s^{\alpha_*-1}}{D_{\alpha_*}}F(x_*,0)\,\frac{2}{k}\Gamma\!\left(\frac{1}{k}\right)\!\left(\frac{k!}{\alpha^{(k)}(x_*)|\ln(\tau s)|}\right)^{\!1/k},
    \end{split}
    \label{eq:asy-IM-k}
    \end{equation}
    where we have used the change of variable $u = \big(\alpha^{(k)}(x_*)|\ln(\tau s)|/k!\big)^{1/k}(x-x_*)$ to evaluate the integral about $x_*$.
    Applying the Tauberian theorem to \eqref{eq:asy-IM-k} yields
    \begin{equation}
        \Psi(t)  \sim \frac{C\,t^{-\alpha_*}}{\big(\ln(t/\tau)\big)^{1/k}},
    \label{eq:psi_interior_k_time}
    \end{equation}
    where $C = \frac{2F(x_*,0)}{D_{\alpha_*}k \Gamma(1-\alpha_*)}\Gamma\!\left(\frac{1}{k}\right)\!\left(\frac{k!}{\alpha^{(k)}(x_*)}\right)^{\!1/k}$.
    
\subsection{Absorbing boundary minimum}
\label{sec:absorbing}
    We now consider when $\alpha(x)$ has a unique minimum $\alpha_*$ on the absorbing boundary at $x  = 0$ such that
    \begin{equation}
        \alpha(x) = \alpha(0)+\alpha'(0)x+ O(x^2),
    \end{equation}
    where $\alpha'(0) > 0$.
    From \eqref{eq:asy-ode}, $F(x,0) = x$ near $x=0$, so \eqref{eq:psi_laplace_method} becomes
    \begin{equation}
    \begin{split}
        \hat\Psi(s) &\sim \frac{s^{\alpha(0)-1}
}{D_{\alpha(0)}}
        \int_{0}^{L}x\,e^{-|\ln(\tau s)|\alpha'(0)x}\,dx \\
        &\sim \frac{s^{\alpha(0)-1}}{D_{\alpha(0)}\alpha'(0)^{2}|\ln(\tau s)|^{2}},
    \end{split}
    \label{eq:psi_left_laplace}
    \end{equation}
    where we have used integration by parts to evaluate the one-sided Laplace integral about $x=0$. 
    Applying the Tauberian theorem to (\ref{eq:psi_left_laplace}) yields
    \begin{equation}
        \Psi(t)\sim\frac{C t^{-\alpha(0)}}{\left(\ln\left( t/\tau \right)\right)^{2}}\,,
    \label{eq:psi_left_time}
    \end{equation}
    where $C=\frac{1} {D_{\alpha(0)}\alpha'(0)^{2}\Gamma(1- \alpha(0))}$.
    If the minimum at $x=0$ is of higher order, with $\alpha^{(j)}(0) = 0$ for $j=1,\dots,k-1$ and $\alpha^{(k)}(0) \neq 0$, a calculation similar to \eqref{eq:psi_interior_k_time} shows that the logarithmic term in \eqref{eq:psi_left_time} has exponent $2/k$.

\subsection{Reflecting boundary minimum}
\label{sec:reflecting}
    Similarly, we consider when $\alpha(x)$ has a unique minimum $\alpha_*=\alpha(L)$ on the reflecting boundary at $x=L$, such that
    \begin{equation}
        \alpha(x)=\alpha(L)+\alpha'(L)(x-L)+O\left((x-L)^2 \right),
    \label{eq:alpha_expand_right}
    \end{equation}
    where $\alpha'(L) < 0$.
    Again from \eqref{eq:asy-ode}, $F(L,0)=x_0$, and \eqref{eq:psi_laplace_method} becomes
    \begin{equation}
    \begin{split}
        \hat\Psi(s) &\sim \frac{1}{D_{\alpha(L)}}s^{\alpha(L)-1}x_0
        \int_{0}^{L}e^{-|\ln(\tau s)||\alpha'(L)|(L-x)}\,dx \\
        &\sim \frac{s^{\alpha(L)-1}x_0}{D_{\alpha(L)}|\alpha'(L)||\ln(\tau s)|},
    \end{split}
    \label{eq:psi_right_laplace}
    \end{equation}
    where we have used integration by parts to evaluate the one-sided Laplace integral about $x=L$. 
    Applying the Tauberian theorem to (\ref{eq:psi_right_laplace}) yields
    \begin{equation}
        \Psi(t)\sim\frac{C\,t^{-\alpha(L)}}{\ln\left( t/\tau \right)},
    \label{eq:psi_right_time}
    \end{equation}
    where $C=\frac{x_0}{D_{\alpha(L)} |\alpha'(L)|\Gamma(1-\alpha(L))}$. If the minimum at $x=L$ is of higher order, with $\alpha^{(j)}(L) = 0$ for $j=1,\dots,k-1$ and $\alpha^{(k)}(L) \neq 0$, a calculation similar to \eqref{eq:psi_interior_k_time} shows that the logarithmic term in \eqref{eq:psi_right_time} is raised to the power of $1/k$.
        
    The survival probabilities when the fractional exponent has a single interior minimum \eqref{eq:psi_interior_time}, multiple interior minima \eqref{eq:psi_multi_time}, a k\textsuperscript{th} order interior minimum \eqref{eq:psi_interior_k_time} and minimum at a boundary, \eqref{eq:psi_left_time} and \eqref{eq:psi_right_time}, all have power law scaling according to the minimum value of the fractional exponent $\alpha_*$.
    Physically, this is intuitive when considering the underlying random walks \cite{fedotov2019asymptotic,angstmann2026global} that lead to \eqref{eq:vofde} in the diffusion and continuum limits.
    As we consider long time asymptotics, the strongest trap dominates in contribution to the FPT statistics.
    Interestingly, the logarithmic correction $(\ln t)^{-\nu}$ is sensitive to the shape of $\alpha(x)$ and $F(x,0)$ near the minimizing point.
    The different scenarios are $\nu=\tfrac{1}{k}$ for k\textsuperscript{th}-order interior minima, $\nu=1$ for minimum at the reflecting boundary, and $\nu=2$ for minimum at the absorbing boundary.
     To make explicit the difference between the variable-order and constant-order cases, we provide the FPT distribution of the fractional diffusion equation \eqref{eq:vofde} when the fractional exponent is constant.
    Setting $\alpha(x) = \alpha$ constant in \eqref{eq:vofde} reduces it to the standard time-fractional diffusion equation
    \begin{equation}
        \partial_t p = D_\alpha\,\mathcal{D}_t^{1-\alpha}\partial_x^2 p.
    \end{equation}
    The solution to this equation given $p(x,0) = \delta(x-x_0)$, absorbing boundary $p(0,t)=0$, and reflecting boundary $\partial_x p(L,t)=0$ is \cite{metzler2000random}
    \begin{equation}
        p(x,t) = \sum_{n=0}^\infty \phi_n(x)\phi_n(x_0)\,E_\alpha\!\left( -D_\alpha k_n^2 t^\alpha\right),
        \label{eq:density_constant}
    \end{equation}
    with $\phi_n(x) = \sqrt{2/L}\sin(k_n x)$ and $k_n = (n+1/2)\pi/L$.
    Integrating \eqref{eq:density_constant} over $[0,L]$ and applying $E_\alpha(-z) \sim 1/[z\Gamma(1-\alpha)]$ as $z \to \infty$ gives
    \begin{equation}
        \begin{split}
            \Psi(t) &\sim \frac{t^{-\alpha}}{D_\alpha \Gamma(1-\alpha)}\left[\sum_{n=0}^\infty \frac{2}{L}\frac{\sin((n+1/2)\pi x_0/L)}{((n+1/2)\pi/L)^3}\right]\\
            &\sim \frac{x_0 L - x_0^2/2}{D_\alpha \Gamma(1-\alpha)}\,t^{-\alpha}.
        \end{split}
        \label{eq:psi_constant_tail}
    \end{equation}
    %
%    The same result can be obtained via \eqref{eq:asy-final} by setting $\alpha(x) = \alpha$ constant.
%    %
%    Substituting $F(x,0) = \min\{x,x_0\}$ into \eqref{eq:asy-final}, gives
%    \begin{equation}
%        \begin{split}
%        \hat\Psi(s) &\sim \frac{2\tau^\alpha}{a^2}\,s^{\alpha-1}\int_0^L \min\{x,x_0\}\,dx\\
%        &= \frac{2\tau^\alpha(x_0 L - x_0^2/2)}{a^2}\,s^{\alpha-1},
%        \end{split}
%    \end{equation}
%    whose Tauberian inversion is equal to \eqref{eq:psi_constant_tail}.
    %
    Comparing \eqref{eq:psi_constant_tail} with \eqref{eq:psi_interior_time}, \eqref{eq:psi_multi_time}, \eqref{eq:psi_interior_k_time}, \eqref{eq:psi_left_time} and \eqref{eq:psi_right_time}, it is clear that variable-order fractional diffusion generates an additional logarithmic correction to the FPT distribution.
 \rev{This result can be intuitively interpreted through the CTRW that underlies \eqref{eq:vofde} \cite{metzler2000random,fedotov2019asymptotic}. 
    The local exponent $\alpha(x)$ fixes the heavy tail of the waiting-time distribution at position $x$, so that the region of smallest $\alpha(x)$ traps walkers with the most slowly decaying waiting times.
    At long times the first-passage statistics are therefore dominated by walkers that have visited this slowest region, which yields the $t^{-\alpha_*}$ power law. 
    The accompanying logarithmic factor, controlled by the local shape of $\alpha(x)$, encodes the spatial measure of that slow region and how the dominant trapping is distributed about the minimum.}

	%%%%%%%%%%%%%%%%%%%%%%%%%%%%%%%%%%%%%%%%%%%%%%%%%%%
	%%%%%%%%%%%%%%%%%%%%%%%%%%%%%%%%%%%%%%%%%%%%%%%%%%%
	
\section{One-dimensional examples}
\label{sec:examples1d}

\subsection{Linear space dependence}
\label{sec:linear}
    When the fractional exponent is a linear function of space, $\alpha(x) = c+bx$, the exact Laplace-space solution exists \cite{roth2020inhomogeneous} and we can calculate the asymptotic FPT distributions directly.
    In \cite{roth2020inhomogeneous}, a uniform initial condition was considered to derive the exact Laplace space solution. 
    Here, we consider the initial condition \eqref{eq:ICFPT}, which is more convenient for FPT calculations.
    The appendix shows the exact solutions and their asymptotic expansions, following the methodology of \cite{roth2020inhomogeneous}.
        
            \begin{figure}
        \centering
        \includegraphics[width=\columnwidth]{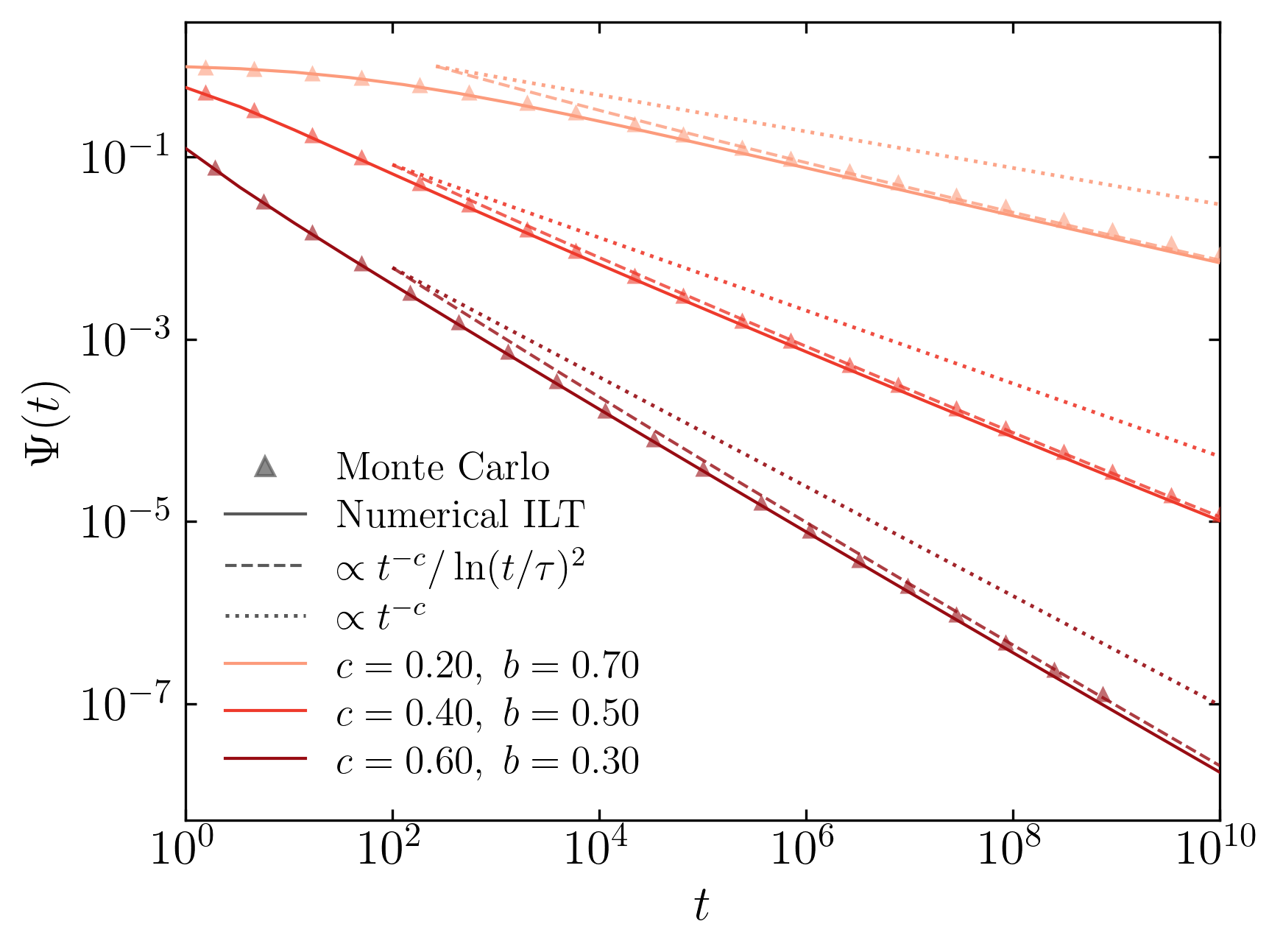}
        \caption{Survival probability $\Psi(t)$ \eqref{eq:psi_bpos_time} for the linear profile $\alpha(x)=c+bx$ with $b>0$. 
        The minimum, $\alpha_*$, is located at the absorbing boundary $x=0$ and initial condition \eqref{eq:ICFPT} was used with $x_0=0.5$. 
        The parameter values used were $L=1$, $\tau=10^{-5}$ and $a=0.02$. 
        The triangles denote Monte Carlo simulations with $N=10^8$ particles performed using methods outlined in \cite{fedotov2019asymptotic}. 
        The solid lines show the numerical inverse Laplace transform of \eqref{eq:psi_exact_boundary} using methods outlined in \cite{kuhlmann2013inverse,roth2020inhomogeneous}. 
        The dashed lines show the asymptotic scaling \eqref{eq:psi_bpos_time}.
        Different colours denote a different set of values $(c,b)$.
        \rev{The dotted lines show the pure power law $t^{-\alpha_*}$ with no logarithmic correction, that is the constant-order result \eqref{eq:psi_constant_tail} at the same $\alpha_*$.
        They are included to show that the logarithmic factor is a qualitative departure from power-law decay and not an adjustable fit.}
        }
    \label{fig:survival_linear_pos}
    \end{figure}
    \begin{figure}
    	\centering
    	\includegraphics[width=\columnwidth]{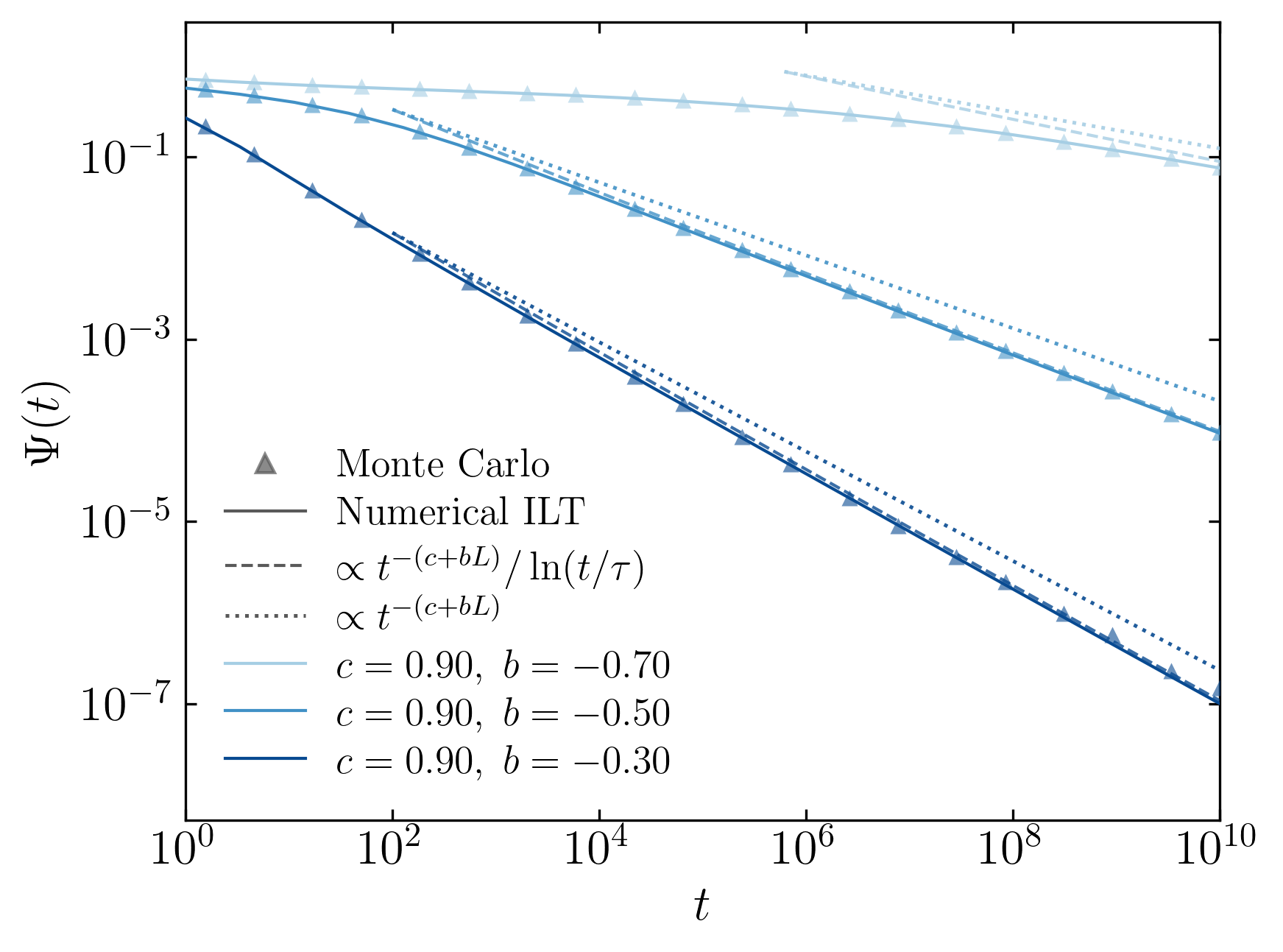}
	\caption{The same as Fig. \ref{fig:survival_linear_pos} but for $b<0$, which places the minimum at the reflecting boundary ($x=L$).
	The dashed lines now show the Tauberian prediction \eqref{eq:psi_bneg_time}.
	\rev{The dotted lines again show the pure power law $t^{-\alpha_*}$ for comparison.}
	The parameters $(c=0.9, b=-0.7)$, $(c=0.9, b=-0.5)$ and $(c=0.9, b=-0.3)$ match $(c=0.2, b=0.7)$, $(c=0.4, b=0.5)$ and $(c=0.6, b=0.3)$ in Fig. \ref{fig:survival_linear_pos} such that $\alpha_* = 0.2$, $\alpha_* = 0.4$ and $\alpha_* = 0.6$ in both plots, respectively.
	}
	\label{fig:survival_linear_neg}
    \end{figure}
    
    When $b>0$, \eqref{eq:psi_left_time} gives, 
    \begin{equation}
	   \Psi(t)\sim\frac{2\tau^c}{a^2 b^2\,\Gamma(1-c)}\,\frac{t^{-c}}{\left(\ln \left(t/\tau \right)\right)^2},
	\label{eq:psi_bpos_time}
    \end{equation}
    and when $b<0$, \eqref{eq:psi_right_time} gives,
    \begin{equation}
	   \Psi(t)\sim\frac{2x_0\tau^{c+bL}}{a^2|b|\,\Gamma(1-c-bL)}\,\frac{t^{-(c+bL)}}{\ln \left( t/\tau \right)}.
	\label{eq:psi_bneg_time}
    \end{equation}
    Figures \ref{fig:survival_linear_pos} and \ref{fig:survival_linear_neg} show excellent correspondence between the asymptotic scalings, \eqref{eq:psi_bpos_time} and \eqref{eq:psi_bneg_time} respectively, Monte Carlo simulations and the numerical inverse Laplace transform of the exact solution in Laplace space \eqref{eq:psi_exact_boundary} (see Appendix).
	%
%    In Figs. \ref{fig:survival_linear_pos} and \ref{fig:survival_linear_neg}, we have ensured that the parameter sets $(c,b)$ between the two plots have the same minimum fractional exponent $\alpha_*$, i.e. for $(0.2, 0.7)$ in Fig. \ref{fig:survival_linear_pos} and $(0.9, -0.7)$ in Fig. \ref{fig:survival_linear_neg} both have $\alpha_*=0.2$ and so on for $\alpha_*=0.4$ and $0.6$. 
    %
    Comparing Figs. \ref{fig:survival_linear_pos} and \ref{fig:survival_linear_neg}, we see that even when the minimum of the fractional exponent, $\alpha_*$ are equal, the location of the minimum causes the logarithmic correction to generate heavier tails when $b<0$.
    This heavier tail is intuitive as the `ultra-slow anomalous advection' generated by the space dependence of the fractional exponent \cite{fedotov2019asymptotic} pushes the random walk away from the absorbing boundary in the case of $b<0$, leading to heavier tails in the FPT survival probability.
    
    %%%%%%%%%%%%%%%%%%%%%%%%%%%%%%%%%%%%%%%%%%%%%%%%%%%
	%%%%%%%%%%%%%%%%%%%%%%%%%%%%%%%%%%%%%%%%%%%%%%%%%%%
	
\subsection{Gaussian wells}
\label{sec:gaussian}
As an extension, we consider a fractional exponent that is a non-linear function of position possessing two minima,
	\begin{equation}
	\alpha(x) = \alpha_0 - A\big(\phi(x-x_1) + \phi(x-x_2)\big),	
	\label{eq:GaussianWells}
	\end{equation}
	with $\phi(x) = \exp(-x^2/(2\sigma^2))$ and $\alpha_0, A\in\mathbb{R}$ being real constants such that $0<\alpha(x)<1$.
    The minimum value $\alpha_* \approx \alpha_0 - A$ is attained at both $x_1$ and $x_2$, provided the two wells are well separated ($\sigma \ll |x_1 - x_2|$). 
    Applying \eqref{eq:psi_multi_time} with $F(x_j,0) = \min\{x_j, x_0\}$, $\alpha_* = \alpha_0 - A$, and $\alpha''(x_j) = A/\sigma^2$ at both wells yields
    \begin{equation}
	\begin{split}
		\Psi(t) \sim\;& \frac{2\tau^{\alpha_*}}{a^2\,\Gamma(1-\alpha_*)}\sqrt{\frac{2\pi\sigma^2}{A}}\\
		&\times\big(\min\{x_1,x_0\} + \min\{x_2,x_0\}\big)\,\frac{t^{-\alpha_*}}{\sqrt{\ln(t/\tau)}}.
	\end{split}
	\label{eq:psi_gaussian_wells}
    \end{equation}
    
\begin{figure}
        \centering
        \includegraphics[width=0.9 \columnwidth]{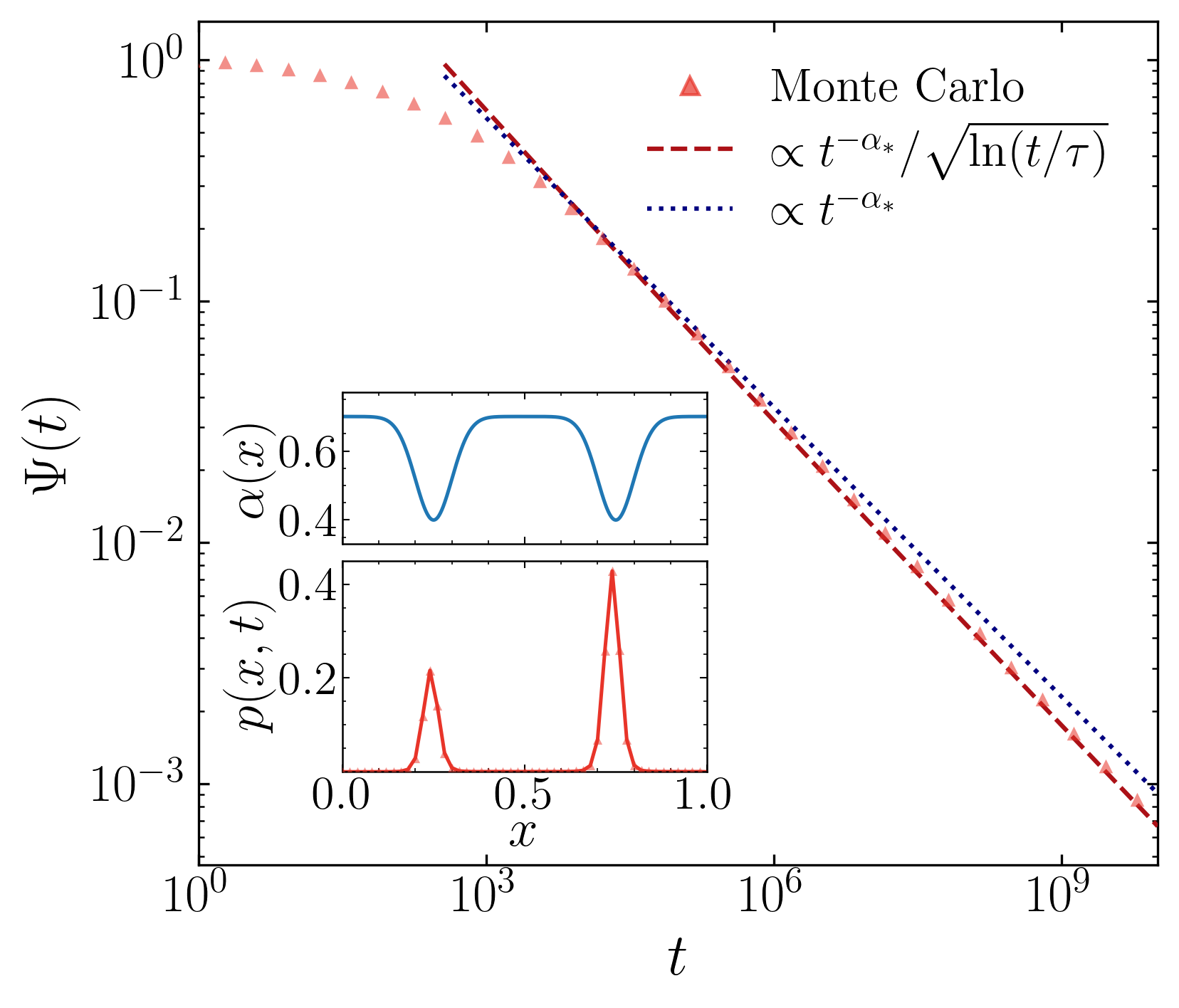}
        \caption{Survival probability $\Psi(t)$ for the Gaussian wells profile \eqref{eq:GaussianWells} with initial condition \eqref{eq:ICFPT} at $x_0=0.5$. 
        The parameter values used were $L=1$, $\tau=10^{-3}$ and $a=0.02$, $x_1 = 0.25$, $x_2=0.75$, $\sigma = 0.05$, $\alpha_0=0.7$ and $A=0.3$.
        The triangles denote Monte Carlo simulations with $N=10^7$ particles performed using methods outlined in \cite{fedotov2019asymptotic}. 
        The dashed line shows the asymptotic scaling \eqref{eq:psi_gaussian_wells}.
        The dotted line shows the pure power law $t^{-\alpha_*}$ with no logarithmic correction.
        \textit{Upper inset}: A plot of \eqref{eq:GaussianWells}.
        \textit{Lower inset}: A plot of the PDF at time $t=10^{6}$.
        }
        \label{fig:survival_gaussian}
    \end{figure}

    Figure \ref{fig:survival_gaussian} shows the excellent correspondence between the asymptotic prediction \eqref{eq:psi_gaussian_wells} and Monte Carlo simulations of the underlying random walk governed by \eqref{eq:vofde} when the fractional exponent varies as \eqref{eq:GaussianWells}.
    The PDF obtained via Monte Carlo in the lower inset of Fig. \ref{fig:survival_gaussian} confirms that the random walk accumulates at the minima of the Gaussian wells, consistent with the physical picture that the deepest traps control the FPT statistics at long times.
 \rev{This accumulation at the minima of the fractional exponent is the `anomalous-aggregation' phenomenon for space-dependent subdiffusion \cite{fedotov2012subdiffusive,fedotov2019asymptotic}. 
    The regions of smallest $\alpha(x)$ act as the strongest traps, dominating the FPT statistics at asymptotically long times.
    The probability density is progressively transferred into these dominating traps, similar to how a zero-temperature system relaxes in an effective potential landscape with $\alpha(x)$ playing the role of the potential. 
    For an interior minimum the long-time particle distribution offers an independent and complementary way to locate the minima of $\alpha(x)$, although the approach to this aggregated state is itself only logarithmically slow in time \cite{fedotov2019asymptotic}.}
        
	The Monte Carlo simulation method used in Figs. \ref{fig:survival_linear_pos}, \ref{fig:survival_linear_neg} and \ref{fig:survival_gaussian} is outlined in \cite{fedotov2019asymptotic} modified with an absorbing boundary at $x=0$.
    For Figs.~\ref{fig:survival_linear_pos}, \ref{fig:survival_linear_neg} and \ref{fig:survival_gaussian}, we also show the pure power law $t^{-\alpha_*}$ (dotted lines) without the logarithmic correction. 
    This is the constant-order result \eqref{eq:psi_constant_tail} evaluated at the minimum $\alpha_*$, and it shows how large the logarithmic correction is over the simulated time windows. 
    The gap between this curve and the asymptotic prediction keeps growing with $t$, so the logarithmic factor is a qualitative departure from power-law decay rather than a correction that could be absorbed into a fitted exponent.
    We outline a method for inference in the next section.
    Another simulation method in continuous space using Brownian increments has also been outlined in \cite{sibatov2024variable}.

\section{Parameter inference from first passage data}
\label{sec:experiment}

    Here we recover the parameters $\alpha_*$ and $\nu$ from FPT data, first by a linear regression that makes the structure of the problem visible.
    We formulate a conditional survival as
    \begin{equation}
        \Psi(t|u) = \frac{\Psi(t)}{\Psi(u)},
        \label{eq:con_survival}
    \end{equation}
    where $t>u$, and the asymptotic survival probability for any twice differentiable $\alpha(x)$ is
    \begin{equation}
        \Psi(t)\sim \, Ct^{-\alpha_*}\left[\ln (t/\tau)\right]^{-\nu}.
        \label{eq:psi_general_tail}
    \end{equation}
    Hence \eqref{eq:con_survival} becomes
    \begin{equation}
        \Psi(t|u) \sim \frac{Ct^{-\alpha_*}\left[\ln (t/\tau)\right]^{-\nu}}{Cu^{-\alpha_*}\left[\ln (u/\tau)\right]^{-\nu}} = \left(\frac{t}{u} \right)^{-\alpha_*}\left( \frac{\ln ( t / \tau ) }{ \ln ( u / \tau) } \right) ^{-\nu}.
        \label{eq:asy_con_survival}
    \end{equation}
    We reformulate \eqref{eq:asy_con_survival} into a linear regression problem,
    \begin{equation}
       Y = \alpha_* + \nu X,
       \label{eq:linear-form}
    \end{equation}
    where $Y = -\ln [\Psi(t|u)]/\ln \left(\tfrac{t}{u} \right)$ and $X = \ln\left[\ln \left( \frac{t}{\tau} \right) /\ln \left(\frac{u}{\tau} \right)  \right] / \ln \left(\frac{t}{u} \right)$.
    The power exponent is the intercept and the logarithmic exponent is the slope, and no amplitude enters either.
    A single cutoff $u$ is fixed, and each retained time $t>u$ contributes one point $(X,Y)$, with the conditional survival replaced by its empirical counterpart $\hat\Psi(t\,|\,u)=n(t)/n(u)$, where $n(t)=\#\{i:T_i>t\}$.
    Pooling several cutoffs would add points but no information, since every pair $(u,t)$ is a function of the same one-dimensional $\hat\Psi$.
    The points are not independent, because $n(t)$ and $n(t')$ are counted from the same walkers, and writing $N=n(u)$ for the number retained,
    \begin{equation}
        \mathrm{Cov}\!\left[\ln\hat\Psi(t),\ln\hat\Psi(t')\right]
        =\frac{1}{n(t\wedge t')}-\frac{1}{N},
        \label{eq:greenwood}
    \end{equation}
    so $\ln\hat\Psi$ is a Brownian motion in the time $1/n$: its variance grows without bound as the record is exhausted, and neighbouring points are almost perfectly correlated.
    Ordinary least squares assumes the points are independent, so it counts the same information many times over.
    We therefore fit \eqref{eq:linear-form} by generalized least squares with the covariance \eqref{eq:greenwood}, at 21 times equally spaced in $\ln(t/u)$ over $1\le\ln(t/u)\le 18.1$, above the cutoff $\ln(u/\tau)=24$.
    The endpoint is placed where 2000 walkers remain, so that 2001 of the $N=3.74\times10^{6}$ retained events survive the last of them, and the standard errors are those of the generalized least squares covariance.
    \begin{figure}[t]
        \includegraphics[width=\columnwidth]{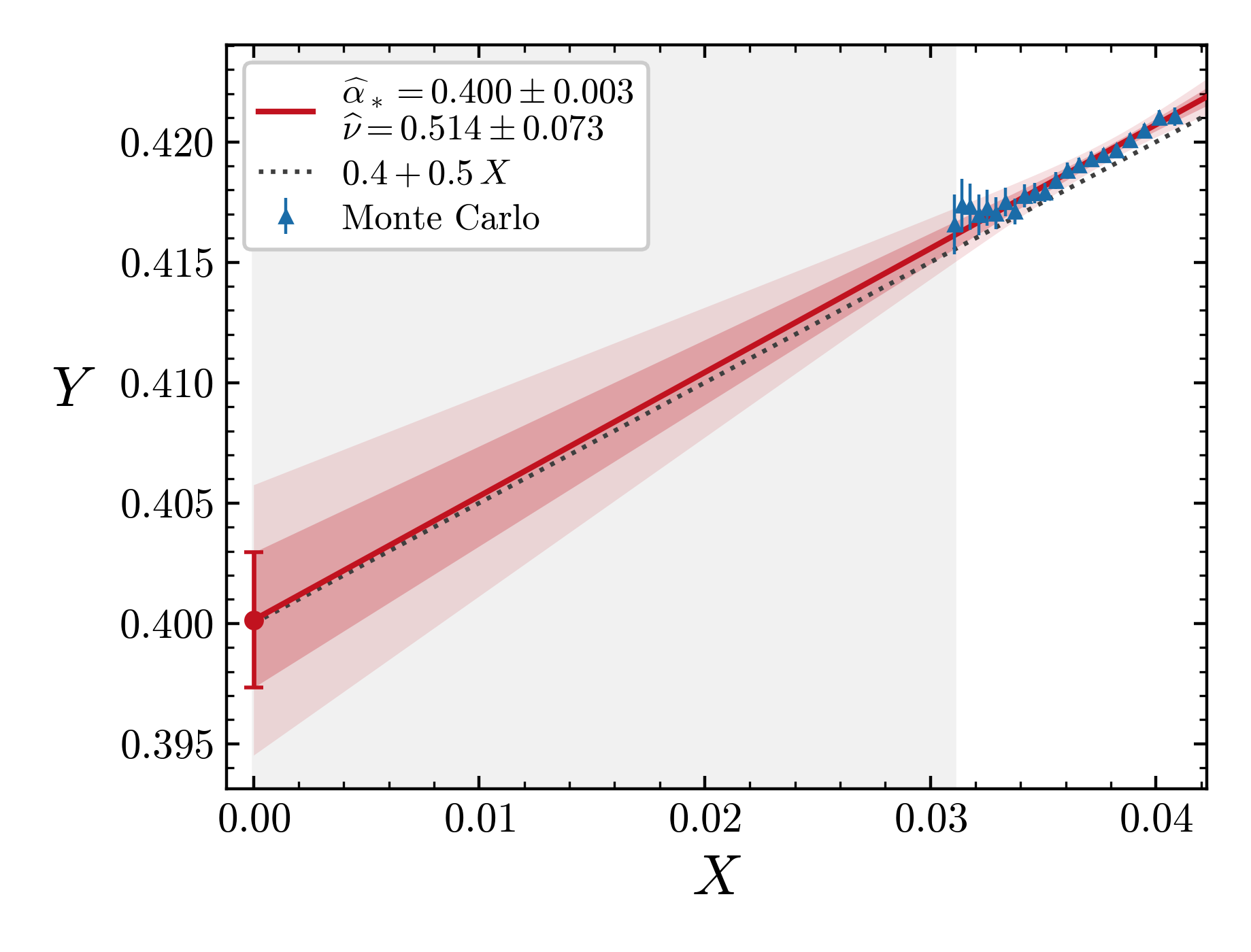}
        \caption{The conditional survival tail \eqref{eq:asy_con_survival} in the linearizing variables of \eqref{eq:linear-form}, at cutoff $\ln(u/\tau)=24$ with $N=3.74\times10^{6}$ retained first passage times.
        Triangles are the empirical conditional survival with standard errors from \eqref{eq:greenwood}, shown where at least 2000 walkers remain.
        The line and its $1\sigma$ and $2\sigma$ bands are the generalized least squares fit, whose intercept is $\alpha_*$ and whose slope is $\nu$.
        The dotted line is the prediction $\alpha_*=0.4$, $\nu=1/2$.
        Shading marks $X$ below the retained range: the intercept lies at $X=0$, always outside the data.}
        \label{fig:tail-regression}
    \end{figure}
    Figure~\ref{fig:tail-regression} shows the regression, giving
    \begin{equation}
        \widehat{\alpha}_*=0.400\pm0.003,
        \qquad
        \widehat{\nu}=0.514\pm0.073 ,
        \label{eq:recovered-gls}
    \end{equation}
    against the true $\alpha_*=0.4$ and the predicted $\nu=1/2$.
    The logarithmic exponent is $7$ standard errors from zero, so the pure power law is excluded, and it lies within $1$ standard error of the predicted value.
    The response in \eqref{eq:linear-form} is built from the survivor counts alone, so $\tau$ reaches the fit only through $\ln(u/\tau)$, and an assumed value shifts that logarithm by a subtraction, $\ln(u/\tau)\to\ln(u/\tau)-\delta$ with $\delta=\ln(\tau_{\text{assumed}}/\tau_{\text{true}})$.
    A misspecification is therefore measured by $\delta$ and not by a ratio of scales: an error of a factor of $\mathrm{e}$ moves $\ln(u/\tau)$ from $24$ to $23$ and shifts $\widehat\alpha_*$ by $0.0007$ and $\widehat\nu$ by $0.038$, $0.24$ and $0.50$ standard errors respectively.
    Because $\tau$ appears only inside that logarithm the bias is suppressed, as $\delta/\ln(u/\tau)$ for $\nu$ and as $\delta/\ln^{2}(u/\tau)$ for $\alpha_*$, so an experiment in which $\tau$ is known only to within an order of magnitude still determines $\nu$ to within its statistical error and $\alpha_*$ essentially independently of $\tau$.
\section{Summary}
\label{sec:summary}
	We have derived the asymptotic FPT distribution for the variable-order time-fractional diffusion equation \eqref{eq:vofde} for general $\alpha(x)$.
	The survival probability universally decays as $\Psi(t) \sim C\,t^{-\alpha_*}/[\ln (t/\tau)]^\nu$, where the power law is set by the global minimum $\alpha_*$ and the logarithmic exponent $\nu$ classifies the local geometry of $\alpha(x)$ at that minimum.
 \rev{The same scaling holds when the generalized diffusion coefficient is a smooth, bounded, positive function of position: carrying $D(x)$ through the reduction leaves the limiting problem for $F(x,0)$ unchanged and replaces the weight in \eqref{eq:psi_laplace_method} by $F(x,0)/D(x)$, so $\alpha_*$ and $\nu$ are unaffected and only the amplitude changes, by a factor $1/D(x_*)$.}
	The logarithmic correction is a qualitative departure from the pure power-law decay \eqref{eq:psi_constant_tail} of constant-order fractional diffusion, and provides a \rev{potential} experimental diagnostic.
		A two-parameter linear regression recovers both exponents from simulated first passage times, with the logarithmic exponent identified within one standard error of its predicted value.
	The asymptotic results were confirmed for linear and Gaussian-well profiles using numerical inversion of exact Laplace-space solutions and Monte Carlo simulations.

 \rev{
    Several limitations of this paper should be noted.
    The results are asymptotic, and the approach to the predicted tail is logarithmically slow, so a substantial separation between the microscopic time scale and the observation window is needed before they apply.
    The boundary results assume $\alpha'(0)\neq0$ at the absorbing end and $\alpha'(L)\neq0$ at the reflecting end, and the interior results assume a first nonvanishing derivative of even order.
    For the medium studied here that cutoff is $u/\tau\simeq3\times10^{10}$, so the separation required between the microscopic time scale and the analysis window is large.
    Finally, it would be interesting to analyse previous experiments, such as \cite{weigel2011ergodic}, that have measured constant $\alpha$ for a continuous process and determine if a logarithmic correction exists.
    }

    \begin{acknowledgments}
    We acknowledge funding from the University of New South Wales Faculty Research Grant and helpful discussions with T. Boecking, J. Gooding, V. Ananthanarayanan, C. Angstmann, B. Henry and B. Huang.
    \end{acknowledgments}

	%%%%%%%%%%%%%%%%%%%%%%%%%%%%%%%%%%%%%%%%%%%%%%%%%%%
	%%%%%%%%%%%%%%%%%%%%%%%%%%%%%%%%%%%%%%%%%%%%%%%%%%% 
	\vspace{-0.28cm}
	
	\appendix
	
	\section{Linear dependence solution and asymptotics}
	
	For $\alpha(x) = c+bx$, the differential equation \eqref{eq:vofde_laplace} can be reduced to an inhomogeneous modified Bessel equation, using $F(x,s)\equiv g(\zeta)$ defined in \eqref{eq:Fs-def}.
    This is achieved via the non-linear transformation \cite{roth2020inhomogeneous}
    	\begin{equation}
		\zeta(x,s)
		=
		2 \sqrt{\omega} \varepsilon^{-1} \mathrm{e}^{\varepsilon x/2},
		\qquad
		\omega = \frac{2 (s\tau)^{c}}{a^{2}},
		\qquad
		\varepsilon = b \ln(s\tau),
		\label{eq:variables}
	\end{equation}
	which converts \eqref{eq:vofde_laplace} to
	    \begin{equation}
        \zeta^2 g''(\zeta) + \zeta g'(\zeta) - \zeta^2 g(\zeta) = -\frac{4}{\varepsilon^2} p(x,0).
        \label{eq:modified_bessel}
    \end{equation}
   
    By considering the homogeneous solution of \eqref{eq:modified_bessel}, we obtain $ F(x,s) = g(\zeta) = C_1 I_0(\zeta) + C_2 K_0(\zeta) + g_p(\zeta)$, where $g_p(\zeta)$ is the particular solution dependent on the initial condition $p(x,0)$. 
    %
%    In fact, \eqref{eq:F_homog_linear} can be rewritten as 
%    \begin{equation}
%	   F(x,s) = C_1(\zeta) I_0(\zeta) + C_2 (\zeta)K_0(\zeta) .
%	\label{eq:F_variation}
%    \end{equation}
    %
    Given the absorbing boundary condition $F(0,s)=0$, the reflecting boundary condition $F'(L,s)=0$ and the initial condition \eqref{eq:ICFPT}, we obtain 
    \begin{equation}
    	F(x,s) = -\frac{1}{W_x[u,v]}
	\begin{cases}
		u(x)\,v(x_0), & 0 \le x \le x_0,\\
		u(x_0)\,v(x), & x_0 \le x \le L,
	\end{cases}
	\label{eq:F_exact}
    \end{equation}
    using variation of parameters.
    In \eqref{eq:F_exact} with $\zeta(0,s) = \zeta_0$ and $\zeta(L,s) = \zeta_L$, $u(x) := K_0(\zeta_0)\,I_0(\zeta) - I_0(\zeta_0)\,K_0(\zeta)$ and $v(x) := K_1(\zeta_L)\,I_0(\zeta) + I_1(\zeta_L)\,K_0(\zeta)$, where $W_x[u,v] = -(\varepsilon/2)\left[K_0(\zeta_0)I_1(\zeta_L) + I_0(\zeta_0)K_1(\zeta_L)\right]$.
    Then in Laplace space, 
%    the exact PDF follows from \eqref{eq:Fs-def} as
%    \begin{equation}
%        \hat p(x,s) = \frac{2}{sa^2}(\tau s)^{c+bx} F(x,s),
%    \end{equation} 
%    and 
    the survival function \eqref{eq:SurvivalProbability} is
    \begin{equation}
	   \hat\Psi(s) = \frac{2}{sa^2}\int_0^L (\tau s)^{c+bx}\,F(x,s)\,dx.
	\label{eq:psi_exact_x}
    \end{equation}
    Evaluating \eqref{eq:psi_exact_x}, we obtain
    \begin{equation}
	\begin{split}
		\hat\Psi(s) = &-\frac{\varepsilon}{2s\,W_x[u,v]}\bigg[\\
		&v(x_0)\Big[K_0(\zeta_0)\zeta I_1(\zeta)
		+ I_0(\zeta_0)\zeta K_1(\zeta)\Big]_{\zeta_0}^{\zeta(x_0,s)}\\
		+\,&u(x_0)\Big[K_1(\zeta_L)\zeta I_1(\zeta)
		- I_1(\zeta_L)\zeta K_1(\zeta)\Big]_{\zeta(x_0,s)}^{\zeta_L}
		\bigg].
	\end{split}
	\label{eq:psi_exact_boundary}
    \end{equation}
	\bibliography{refs}

\end{document}